\documentclass[twocolumn, trackchanges]{aastex701}

\newcommand{\HI}{\ion{H}{1}}

\newcommand{\kms}{\mbox{km~s$^{-1}$}}
\newcommand{\Msol}{\mbox{$M_\odot$}}
\newcommand{\surm}{\mbox{$M_\odot$ pc$^{-2}$}}

\newcommand{\siggas}{\mbox{$\Sigma_{\rm gas}$}}
\newcommand{\sigsfr}{\mbox{$\Sigma_{\rm SFR}$}}
\newcommand{\sigstar}{\mbox{$\Sigma_{\star}$}}

\newcommand{\sightwo}{\mbox{$\Sigma_{\rm H_2}$}}

\newcommand{\coj}{\mbox{$^{12}$CO ($J=1\rightarrow0$)}}

\newcommand{\um}{\mbox{$\micron$}}
\newcommand{\ac}{\mbox{$\arcsec$}}

\newcommand{\Htwo}{\mbox{H$_2$}}

\usepackage{indentfirst}

\shorttitle{Molecular Gas and Star Formation in Dwarf Galaxies}
\shortauthors{Yim and Rey}
\received{August 19, 2025}
\accepted{November 15, 2025}

\begin{document}

\title{Molecular Gas and Star Formation in Dwarf Galaxies Observed by the Atacama Large Millimeter/submillimeter Array}

\author[0000-0002-3426-5854]{Kijeong Yim}
\affiliation{Department of Astronomy and Space Science, Chungnam National University, Daejeon 34134, Republic of Korea}
\email[show]{kijeong.yim@gmail.com}  

\author[0000-0002-0041-6490]{Soo-Chang Rey}
\affiliation{Department of Astronomy and Space Science, Chungnam National University, Daejeon 34134, Republic of Korea}
\email[show]{screy@cnu.ac.kr}




\begin{abstract}

We present a spatially resolved analysis of the molecular star formation law (SFL) and gravitational instability in a sample of nearby dwarf galaxies (NGC 1035, NGC 4310, NGC 4451, NGC 4701, NGC 5692, and NGC 6106), using high-resolution \coj\ data from the Atacama Large Millimeter/submillimeter Array. We estimate the star formation rate (SFR) by combining the Galaxy Evolution Explorer near-ultraviolet and the Wide-field Infrared Survey Explorer 12 \um\ imaging data to examine the relationship between molecular gas and SFR densities on scales of several hundred parsecs. 
We find that the power-law slope of the molecular SFL ranges from 0.62 to 1.08, with an average value of N$=0.81\pm0.18$, increasing to N$=0.87\pm0.05$ when excluding galaxies with poorly constrained CO data. These results are roughly consistent with values observed in massive spiral galaxies, suggesting a universal molecular SFL when analyzed with sufficient resolution and sensitivity. Radial profiles of the Toomre $Q$ parameter remain close to unity across the disks, with minimal radial variation, consistent with a self-regulated star formation model. Our results suggest that, despite their lower mass and metallicity, star formation in dwarf galaxies is governed by the same fundamental physical processes as in larger systems. This highlights the significance of high-resolution molecular gas observations in low-mass galaxies.

\end{abstract}


\section{Introduction} \label{sec:intro}
Dwarf galaxies, characterized by their shallow gravitational potential wells, are particularly sensitive to environmental processes such as tidal interactions (\citealt{2001ApJ...559..754M}, \citeyear{2006MNRAS.369.1021M}), ram-pressure stripping (\citealt{1972ApJ...176....1G}; \citealt{2003AJ....125.1926G}; \citealt{2008ApJ...674..742B}), and feedback mechanisms from supernovae and stellar winds (\citealt{1986ApJ...303...39D}; \citealt{1999ApJ...513..142M}; \citealt{2010Natur.463..203G}). Consequently, they serve as excellent laboratories for understanding the impact of environmental processes on star formation (\citealt{2009ARA&A..47..371T}; \citealt{2015A&A...573A..48P}).

Investigating the star formation law (SFL), which describes the empirical relation between the star formation rate (SFR) and the interstellar gas, in dwarf galaxies provides essential insights into galaxy evolution.
Numerous studies focusing on massive spiral galaxies have established a robust power-law relation between the surface densities of molecular gas (\sightwo) and SFR (\sigsfr), with power-law indices in the range of N $\approx$ 0.8--1.2 (\citealt{2002ApJ...569..157W}; \citealt{2008AJ....136.2846B}; \citealt{2011AJ....142...37S}; \citealt{Sun_2023}). This empirical relation underpins our current theoretical understanding of star formation processes in galaxy evolution models and cosmological simulations (\citealt{2003MNRAS.339..289S}; \citealt{2014Natur.509..177V}; \citealt{2015MNRAS.446..521S}; \citealt{2018MNRAS.473.4077P}).

However, it remains uncertain whether this scaling relation holds in low-mass, low-metallicity dwarf galaxies. Observational studies have long been hampered by the challenges of detecting molecular gas in such faint, low-mass galaxies (e.g., \citealt{1998AJ....116.2746T}; \citealt{2005ApJ...625..763L}; \citealt{2012AJ....143..138S}; \citealt{2013ARA&A..51..207B}). Earlier work was often limited by low resolution and sensitivity. For example, \citet{2005ApJ...625..763L}  detected CO emission primarily in the central regions of 44 dwarf galaxies using the Arizona Radio Observatory telescope, which has a large beam size of 55\ac, thereby restricting spatial resolution. \citet{2012AJ....143..138S} conducted higher-resolution observations (13\ac) using the IRAM 30m telescope, but detected CO in only five out of sixteen dwarf galaxies, necessitating the application of stacking techniques to investigate the molecular SFL. They suggested that dwarf galaxies do not follow the classical Kennicutt-Schmidt (KS) power-law trend observed in massive spirals, potentially reflecting different star formation mechanisms or efficiencies (e.g., \citealt{2012AJ....143..138S}; \citealt{2019ApJ...872...16D}).

To better understand the nature of star formation in dwarf galaxies, it is essential to employ observational data with high resolution and sensitivity. Recent advances in interferometric radio astronomy--particularly through observations from the Atacama Large Millimeter/submillimeter Array (ALMA)--now offer unprecedented opportunities to map the molecular gas distribution in dwarf galaxies with high accuracy (e.g., \citealt{2015Natur.525..218R}; \citealt{2022MNRAS.512.1012C}). The superior capabilities of ALMA significantly improve the quality and reliability of molecular gas measurements, overcoming the limitations faced by earlier studies (e.g., \citealt{Leroy_2021}; \citealt{Sun_2023}).

In this paper, we present a comprehensive investigation of the molecular gas content and its associated SFL in dwarf galaxies using high-quality ALMA CO data. By employing a carefully selected sample from \citet{2022MNRAS.512.1012C}, we aim to robustly characterize the relationship between molecular gas and SFR, explicitly testing whether dwarf galaxies follow or deviate from the well-established molecular SFL observed in massive spiral galaxies (e.g., \citealt{2008AJ....136.2846B}; \citealt{Leroy_2013}). Furthermore, we derive the gravitational instability parameter $Q$ (\citealt{1964ApJ...139.1217T}; \citealt{2011MNRAS.416.1191R}) to assess its role in regulating star formation within these lower-mass systems (e.g., \citealt{2008AJ....136.2782L}; \citealt{2013MNRAS.433.1389R}).

This paper is organized as follows. Section \ref{sec:data} describes the sample selection, the ALMA CO observations and reduction, and the infrared and ultraviolet imaging data. Section \ref{sec:results} presents the radial distributions of molecular gas, stars, and SFR, and discusses the molecular SFL as well as the role of the gravitational instability parameter $Q$ in our dwarf galaxy sample. Finally, Section \ref{sum} summarizes our main findings and conclusions.

\begin{table*}[!bt]
\begin{center}
\caption{Galaxy Properties\label{table:galprop}}
\begin{tabular}{cccccccccc}
\tableline\tableline
Galaxy &R.A. (J2000)& Decl. (J2000) &Morphology  &PA& Inclination& Distance& log$M_*$&R$_{25}$& $M_B$\\
&(h m s)&(\degr\ \arcmin\ \arcsec)&&(\degr)&(\degr)& (Mpc)& ($M_\odot$)&($\arcmin$)&(mag)\\
(1)&(2)&(3)&(4)&(5)&(6)&(7)&(8)&(9)&(10)\\
\tableline
NGC 1035&02 39 29.1&$-$08 07 58.6&SAc &148&79&15.9&9.5&1.04&-18.3 \\
NGC 4310&12 22 26.3&$+$29 12 31.2 &SAB &158&71&16.1&9.3&0.97&-17.4 \\
NGC 4451&12 28 40.6&$+$09 15 32.1 &Sbc &158&49&25.9&9.7&0.67&-17.7\\
NGC 4701&12 49 11.6&$+$03 23 19.4 &SAcd &45&49&16.7&9.4&0.87&-18.4\\
NGC 5692&14 38 18.1&$+$03 24 36.8 &S &38&53&25.4&9.3&0.48&-19.0\\
NGC 6106&16 18 47.2&$+$07 24 39.2 &SAc&141&59&24.3&9.7&1.09&-19.2\\
\tableline
\end{tabular}
\end{center}
\tablecomments{Columns: (1) galaxy name; (2) and (3) galaxy coordinates; (4) morphological type of the galaxies from the NASA/IPAC Extragalactic Database (NED\footnote {http://ned.ipac.caltech.edu}); (5) position angle obtained from the $Spitzer$ 3.6 \um\ or WISE 3.4 \um\ (only for NGC 5692) images using the MIRIAD task \texttt{IMFIT}; (6), (7), and (8) inclination, distance, and the total stellar mass from \citet{2022MNRAS.512.1012C}; (9) optical diameter in arcminutes from HyperLEDA\footnote {http://leda.univ-lyon1.fr} \citep{2014A&A...570A..13M}; (10) absolute $B$ magnitude from \citet{2017ApJ...843...37T}. 
}
\end{table*}

\section{Data and Reduction} \label{sec:data}

\begin{figure*}
\begin{center}
\includegraphics[width=1\textwidth,angle=0]{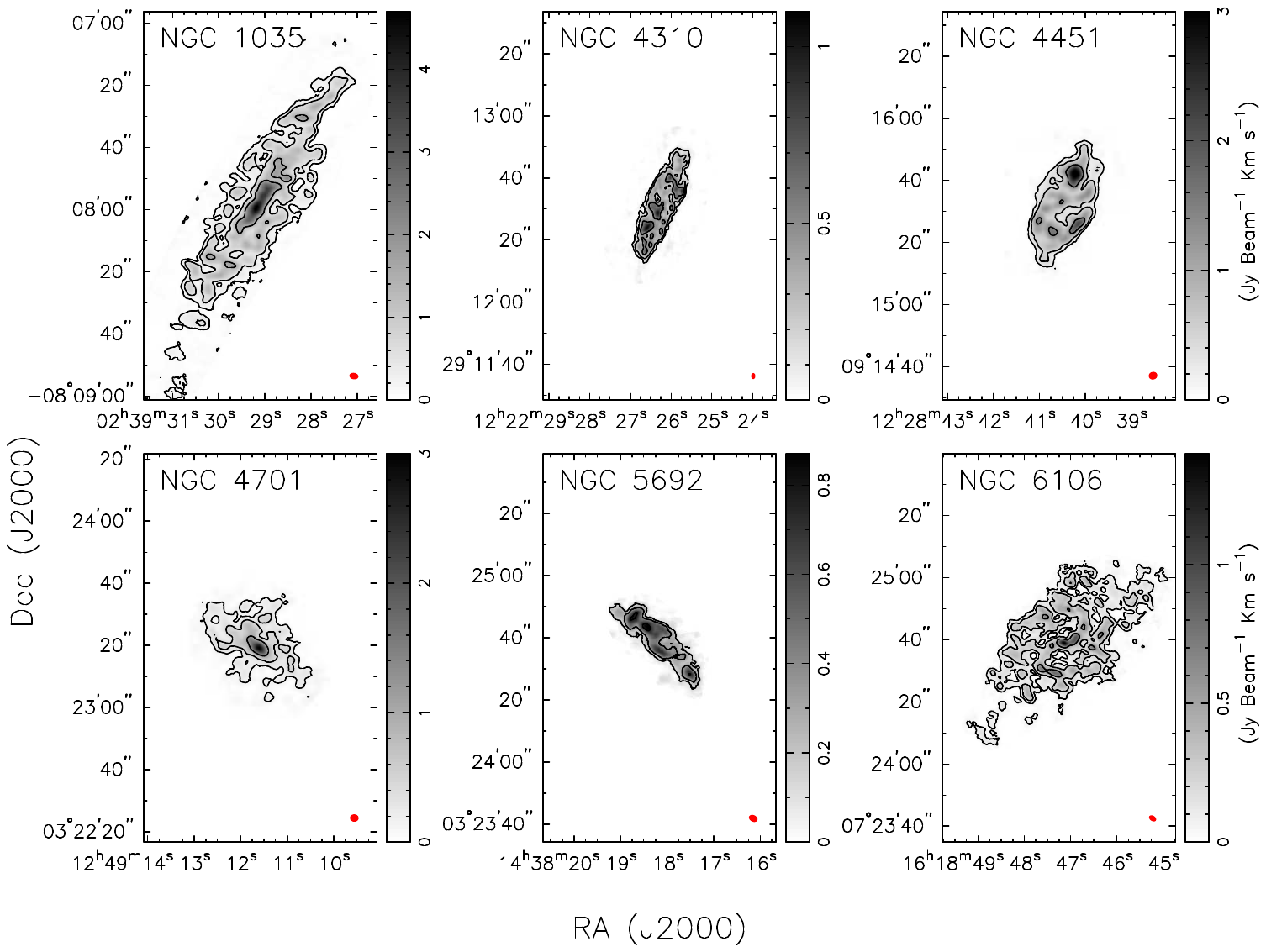}\\
\caption{CO integrated intensity maps for the dwarf galaxies in our sample. Contour levels begin at approximately 5$\sigma$ to clearly delineate reliable molecular gas detections. The synthesized beam size is indicated by the ellipse in the lower-right corner of each panel. 
\label{fig:comap}}
\end{center}
\end{figure*}

\begin{figure*}
\begin{center}
\includegraphics[width=1\textwidth,angle=0]{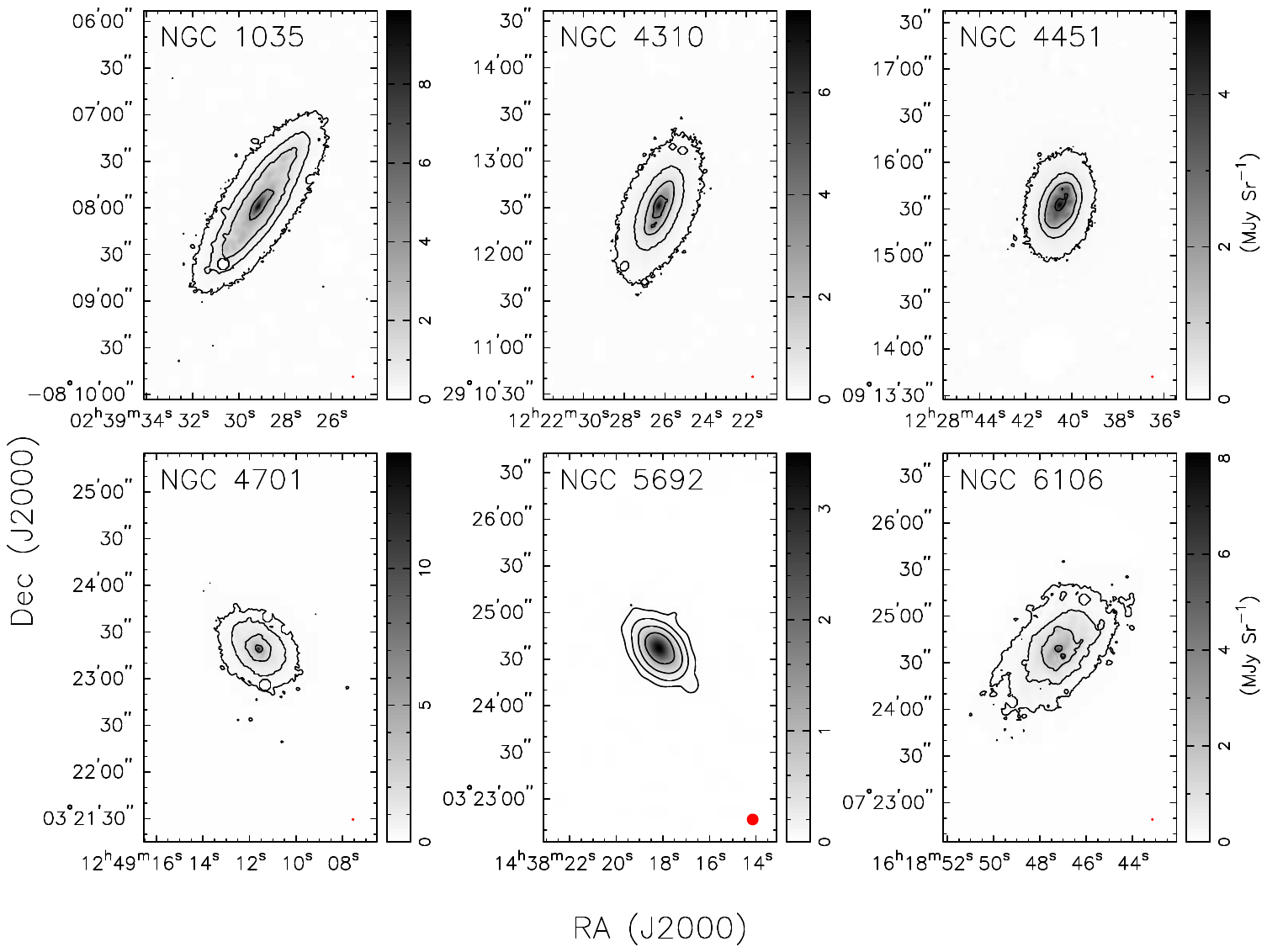}\\
\caption{NIR images used to trace stellar mass distributions. For NGC 5692, the WISE 3.4 \um\ image is shown, while $Spitzer$ IRAC 3.6 \um\ images are displayed for all other galaxies. The contours represent regions of significant stellar emission, beginning at approximately 15$\sigma$. 
The point spread function of the images is marked by the ellipse in the lower-right corner of each panel, illustrating the differences in spatial resolution between the WISE (7$\farcs$5) and $Spitzer$ (1$\farcs$6) images.
\label{fig:stellarmap}}
\end{center}
\end{figure*}

\begin{figure*}
\begin{center}
\includegraphics[width=1\textwidth,angle=0]{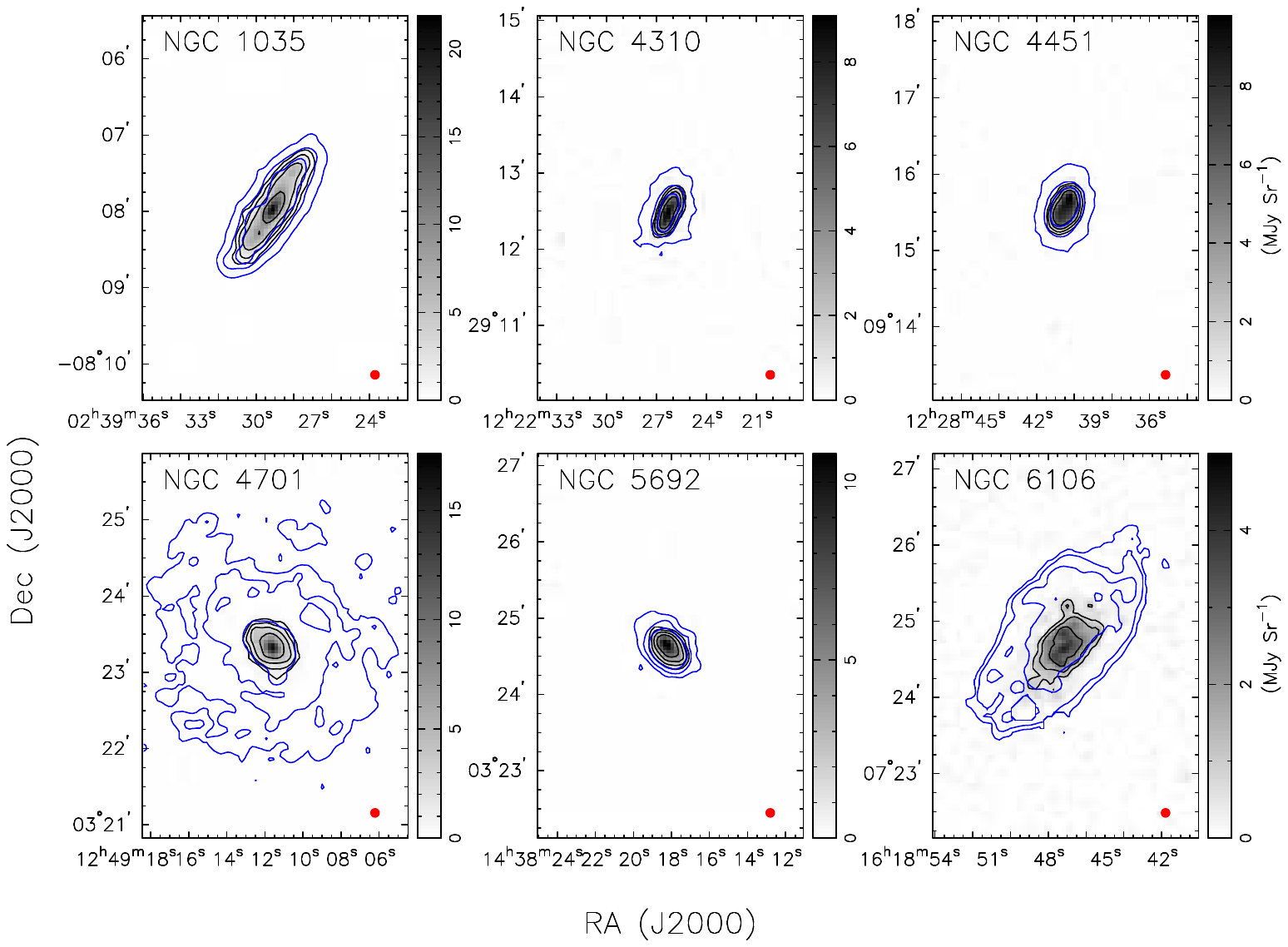}\\
\caption{WISE 12 \um\ images and contours (black) for our dwarf galaxy sample, overlaid with contours from the GALEX NUV emission (blue). The lowest contour level for  both 12 \um\ and NUV emissions is approximately 10$\sigma$. The PSF, with a Gaussian FWHM of $7\farcs5$, is makred by the ellipse in the lower-right corner of each panel. 
\label{fig:sfrmap}}
\end{center}
\end{figure*}

\subsection{Galaxy Sample} \label{sec:sample}
We selected six dwarf galaxies (NGC 1035, NGC 4310, NGC 4451, NGC 4701, NGC 5692, and NGC 6106) from the ALMA Science Archive\footnote {https://almascience.nrao.edu/aq/} (Project ID: 2015.1.00820.S; PI: L. Blitz) originally presented by \citet{2022MNRAS.512.1012C}. Their sample was compiled from the Dwarf Galaxy Dark Matter (DGDM) survey \citep{2017ApJ...843...37T}. The initial DGDM sample comprised 26 dwarf galaxies, selected based on the following criteria: neutral hydrogen (\HI) linewidth $\leq$ 200 \kms, absolute B-band magnitude $(M_B) \geq -18$, optical diameters $>$ 1 arcmin, and strong mid-infrared (MIR) flux. 
The galaxies in our sample are mostly classified morphologically as spirals, with distances $<$ 26 Mpc, stellar masses log$(M_*/M_\odot) < 9.7$, and absolute $B$ magnitudes ranging from -17.4 to -19.2. The detailed properties of these galaxies are listed in Table \ref{table:galprop}.

\subsection{ALMA CO Data} \label{sec:data-co}
We retrieved raw data of \coj\  from the ALMA archive. These observations provide the high sensitivity and spatial resolution needed to detect and analyze faint molecular gas in dwarf galaxies.
Calibration was carried out in the Common Astronomy Software Applications (CASA) package \citep{2007ASPC..376..127M} using scripts provided by ALMA.
The calibrated data were then imaged using the CASA task \texttt{TCLEAN} with Briggs weighting (robust = 0.5) to achieve an effective compromise between sensitivity and spatial resolution. Primary beam corrections were subsequently applied using the task \texttt{IMPBCOR}.  To increase the signal-to-noise ratio, we masked the maps by blanking regions below a 3$\sigma$ threshold in a map smoothed to 6\ac\ resolution.
The resulting integrated intensity maps for our dwarf galaxies are shown in Figure \ref{fig:comap}, highlighting areas of concentrated molecular gas. 
These high-resolution maps allow for detailed spatial analysis of molecular gas distributions, which is crucial for investigating the molecular SFL in dwarf galaxies.
Table \ref{table:obs} summarizes the observational parameters of the ALMA CO maps, including angular resolutions ranging between 1.33\ac\ and 2.93\ac, velocity resolutions of 2 \kms, total fluxes, channel rms noise levels, and systemic velocities (V$_{\rm sys}$). 

\begin{table*}[!bt]
\begin{center}
\caption{CO Observations from ALMA\label{table:obs}}
\begin{tabular}{cccccc}
\tableline\tableline
Galaxy& Angular resolution  & Velocity resolution & Total flux  &Channel noise& V$_{\rm sys}$ \\
&(\ac)&(\kms) &(Jy \kms)&(mK)&(\kms)\\
\tableline
NGC 1035&2.93$\times$2.06& 2& 182.6& 38& 1229 \\
NGC 4310&1.92$\times$1.33& 2& 44.8& 72& 919 \\
NGC 4451&2.76$\times$2.46& 2& 57.6& 54& 868\\
NGC 4701&2.77$\times$2.39& 2& 36.2& 42& 725 \\
NGC 5692&2.89$\times$2.10& 2& 17.6& 45& 1573\\
NGC 6106&2.49$\times$1.68& 2& 63.7& 44& 1464 \\
\tableline
\end{tabular}
\end{center}
\tablecomments{V$_{\rm sys}$ is the systemic velocity obtained from \citet{2017ApJ...843...37T}.}
\end{table*}

\subsection{Near-Infrared 3.4 \um\ and 3.6 \um\ Imaging Data} 
\label{sec:data-stars}

To derive the stellar density distribution, we utilized archival near-infrared (NIR) imaging data from two sources. The Wide-field Infrared Survey Explorer (WISE) 3.4 µm image for NGC 5692 was retrieved from the $z = 0$  Multiwavelength Galaxy Synthesis ($z$0MGS\footnote {https://irsa.ipac.caltech.edu/data/WISE/z0MGS}) project \citep{Leroy_2019}, while $Spitzer$ IRAC 3.6 µm images for all other galaxies were obtained from the Spitzer Survey of Stellar Structure in Galaxies (S$^4$G; \citealt{2010PASP..122.1397S}).

We masked foreground bright stars in the $Spitzer$ 3.6 \um\ images using the mask map provided by S$^4$G. For the WISE 3.4 \um\ image, masking was performed using the Groningen Image Processing System (GIPSY; \citealt{1992ASPC...25..131V}) task \texttt{BLOT}. The point-spread function (PSF) FWHM for the Spitzer 3.6 \um\ images is approximately 1$\farcs$6, while the WISE 3.4 \um\ image from $z$0MGS \citep{Leroy_2019} has been convolved to a Gaussian PSF FWHM of 7$\farcs$5. Figure \ref{fig:stellarmap} shows the NIR images of our sample galaxies, clearly displaying the stellar mass distributions. The contour levels indicate regions of significant stellar emission, with the lowest contour corresponding to approximately 15$\sigma$, highlighting well-defined stellar disks across all galaxies.

\subsection{12 \um\ and Near-Ultraviolet Imaging Data} \label{sec:data-sfr}

We obtained archival 12 \um\ MIR images from WISE and near-ultraviolet (NUV) images from the Galaxy Evolution Explorer (GALEX), both of which are provided by the $z$0MGS project archive. The MIR emission at 12 \um\ primarily traces dust that is heated by young massive stars, with a timescale of $\sim10$ Myr (\citealt{2008AJ....136.2782L} and references therein). Meanwhile, the NUV emission represents star formation over a longer timescale, extending up to $\sim$100 Myr (\citealt{1998ApJ...498..541K}; \citealt{2013Ap&SS.343..361P}). \citet{Leroy_2019} convolved both sets of images to a common Gaussian PSF of 7$\farcs$5, using kernels from \citet{2011PASP..123.1218A}. We also masked bright foreground stars in these images using the GIPSY task \texttt{BLOT}. 

Figure \ref{fig:sfrmap} illustrates the WISE 12 \um\ images and contours (black), overlaid with GALEX NUV contours (blue), which clearly indicate regions of active star formation. The lowest contour level for both the 12 \um\ and NUV images corresponds to approximately 10$\sigma$: $\sim$0.8 MJy sr$^{-1}$ for 12 \um\ and $\sim$0.002 MJy sr$^{-1}$ for NUV. Notably, the spatial extent of NUV emission is significantly larger than that of the corresponding 12 \um\ emission in all galaxies (particularly NGC 4701 and NGC 6106), consistent with a picture in which dust is more centrally concentrated than young stellar populations in dwarf galaxies (e.g., \citealt{2009ApJ...703..517D}).

\section{RESULTS AND DISCUSSION}\label{sec:results}
\subsection{Radial Distributions of Molecular Gas, Stars, and SFR} \label{sec:distribution}

To investigate the relationship between molecular gas and star formation activity in dwarf galaxies, we first derived the radial surface density profiles for three key components: molecular gas (\sightwo), SFR (\sigsfr), and stellar mass (\sigstar). These profiles serve as the basis for our subsequent analysis of the molecular SFL (Section \ref{sec:sflaw}) and gravitational instability (Section \ref{sec:gravi}).

For consistency in spatial resolution across all datasets, we convolved the CO data cubes and the 3.6 \um\ images to the common 7.5\ac\ Gaussian PSF, matching that of the WISE 3.4 \um, WISE 12 \um, and GALEX NUV images, which had already been smoothed to this resolution in the $z$0MGS archive \citep{Leroy_2019}. 
The convolution was carried out using the MIRIAD task \texttt{CONVOL} for CO data and Gaussian kernels from  \citet{2011PASP..123.1218A} for 3.6 µm images.

The \Htwo\ surface mass density was derived from the intensity of CO ($I_{\rm CO}$) using the standard Galactic CO-to-\Htwo\ conversion factor $X_{\rm CO}= 2 \times10^{20} \rm cm^{-2}\,  [K \,\kms]^{-1}$ (\citealt{1996A&A...308L..21S}; \citealt{2001ApJ...547..792D}):
\begin{equation}
\sightwo \, [\surm] = 1.36\, {\rm \cos\,} i \times 3.2 I_{\rm CO} \,[\rm K \,\kms],
\label{eq:sightwo}
\end{equation}
where the factor of 1.36 accounts for the helium correction. Additionally, an inclination correction of $\cos i$ was applied for inclined galaxies, using the inclination values from Table \ref{table:galprop}. The adopted conversion factor $X_{\rm CO}$ corresponds to $\alpha_{\rm CO} = 4.35\ M_\odot\ \rm pc^{-2}\ [K \,\kms]^{-1}$, including the helium correction. 

The SFR surface density 
was estimated using a linear combination of NUV and 12 \um\ intensities given by \citet{2020ApJ...892..148S}, 
following the empirical calibration of \citet{Leroy_2019}:
\begin{eqnarray}
\sigsfr [\Msol \, {\rm kpc^{-2} \, yr^{-1}}] = (8.9 \times 10^{-2}\, I_{\rm NUV} \nonumber\\
+ 4.1 \times 10^{-3}\, I_{12\, \mu m})\, {\rm cos}\ i,
\label{eq:sigsfr}
\end{eqnarray}
where both the NUV and 12 \um\ intensities are measured in units of MJy sr$^{-1}$.

The stellar mass density was calculated from the 3.6 \um\ or 3.4 \um\ maps using the calibration from \citet{Leroy_2021}, assuming a constant mass-to-light ratio of $\Upsilon_\star^{3.4}= 0.35\ \Msol L_{\odot}^{-1}$, which is appropriate for star-forming galaxies \citep{Leroy_2019}:
\begin{eqnarray}
\frac{\sigstar}{\surm}=350\, \cos\,i \left(\frac{\Upsilon_\star^{3.4}}{0.5}\right)\,\frac{I\rm_{3.6\, \mu m}}{\rm MJy \,\,sr^{-1}},\\
\frac{\sigstar}{\surm}=330\, \cos\,i \left(\frac{\Upsilon_\star^{3.4}}{0.5}\right)\,\frac{I\rm_{3.4\, \mu m}}{\rm MJy \,\,sr^{-1}}.
\label{eq:sigstar}
\end{eqnarray}
Although the mass-to-light ratio is known to vary with age or metallicity, adopting a constant value is a reasonable first-order assumption for disk galaxies \citep{2014AJ....148...77M}. 

\begin{figure*}
\begin{center}
\begin{tabular}{c@{\hspace{0.1in}}c@{\hspace{0.1in}}c}
\includegraphics[width=0.3\textwidth]{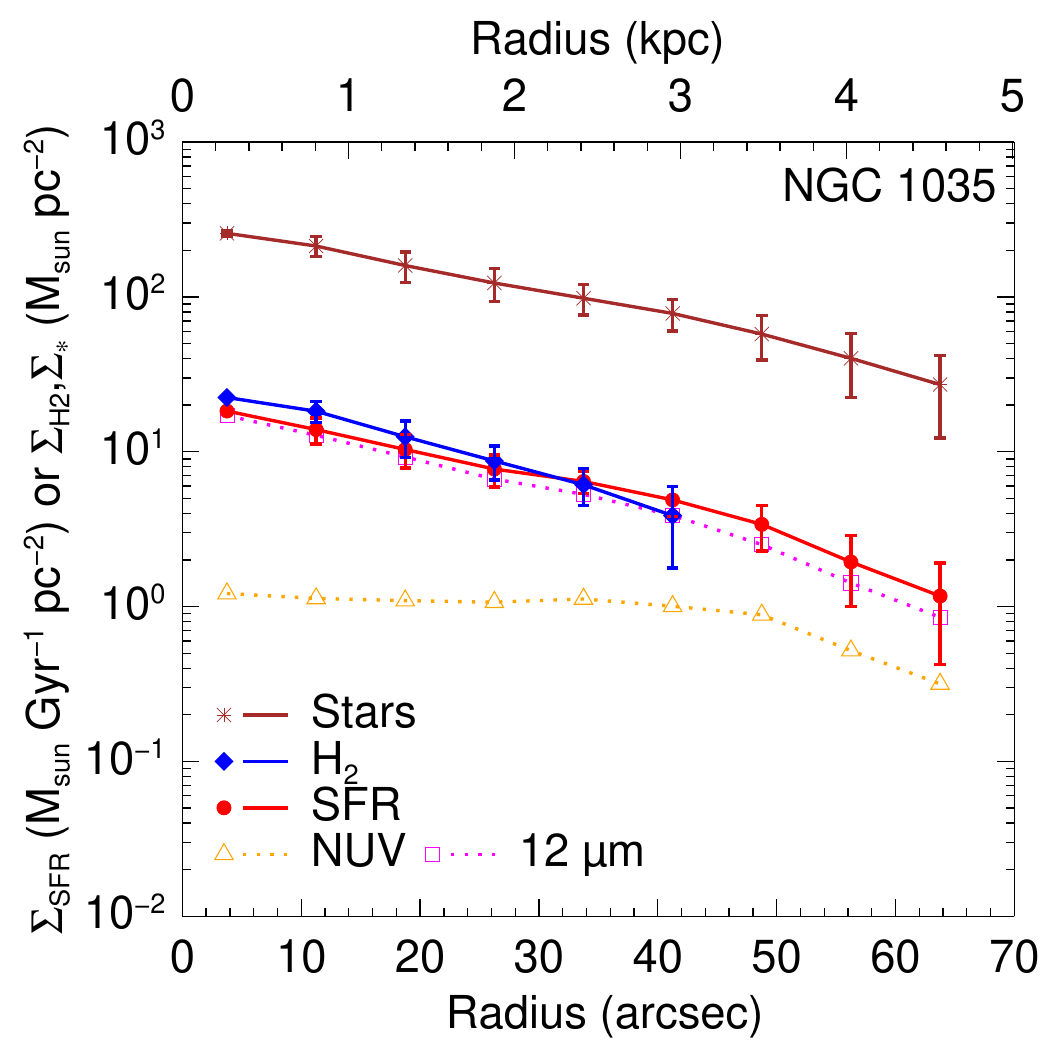} 
\includegraphics[width=0.3\textwidth]{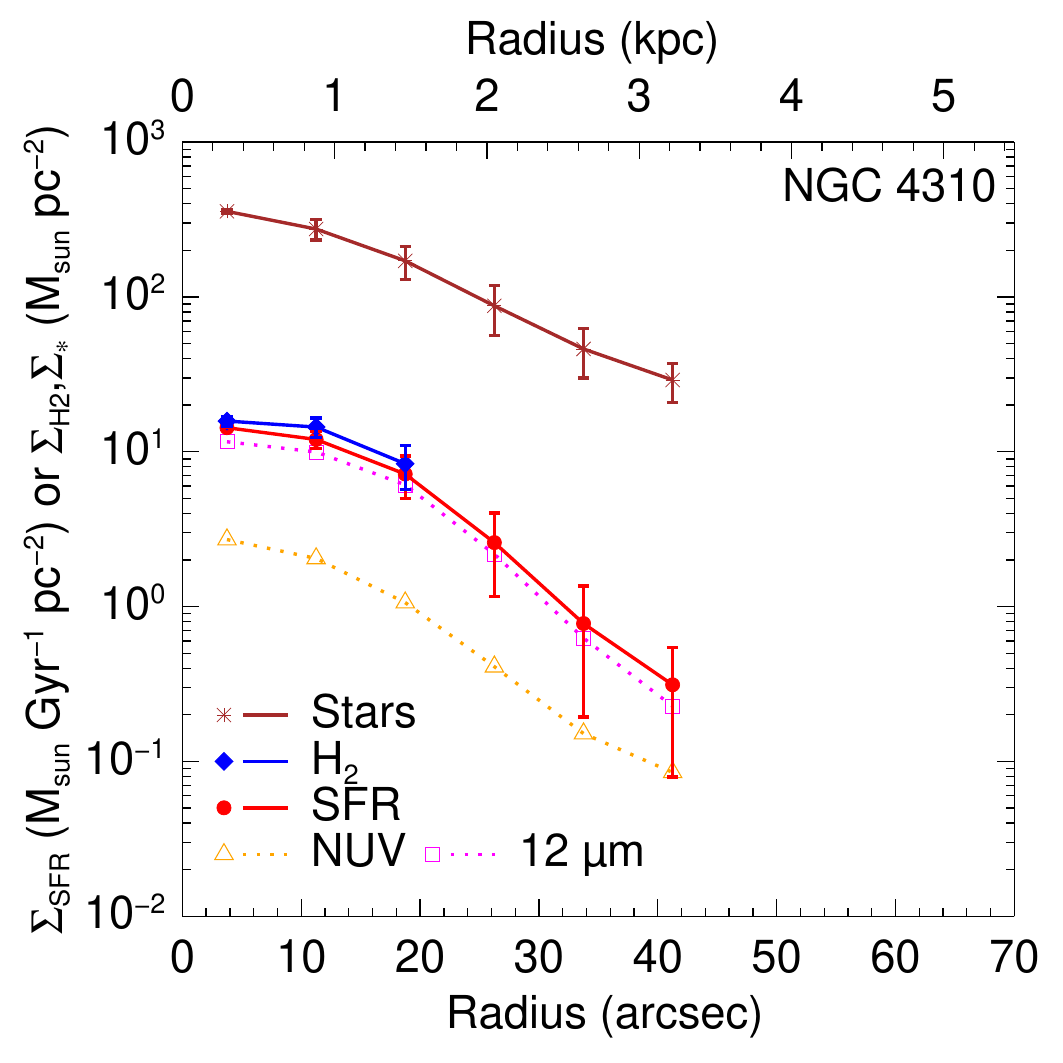}
\includegraphics[width=0.3\textwidth]{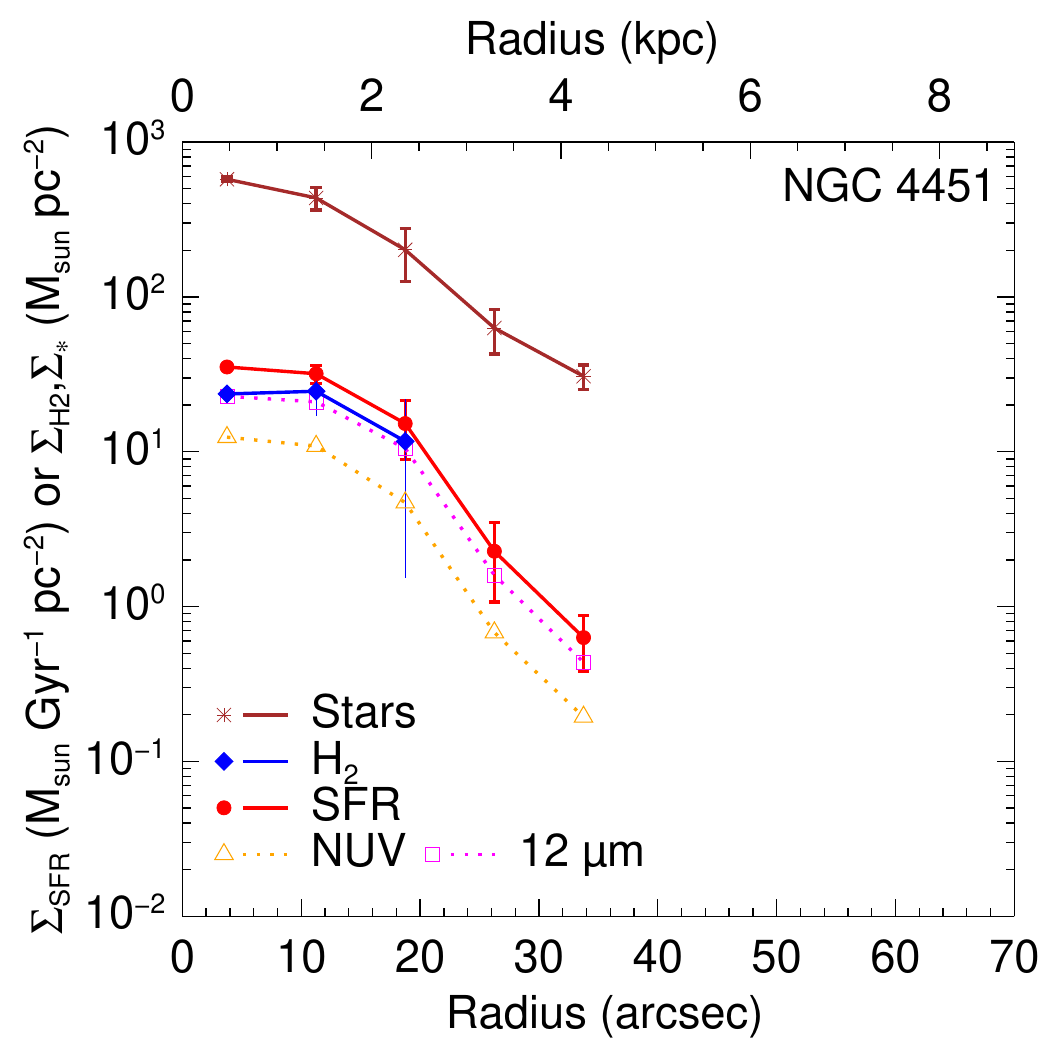}\\
\includegraphics[width=0.3\textwidth]{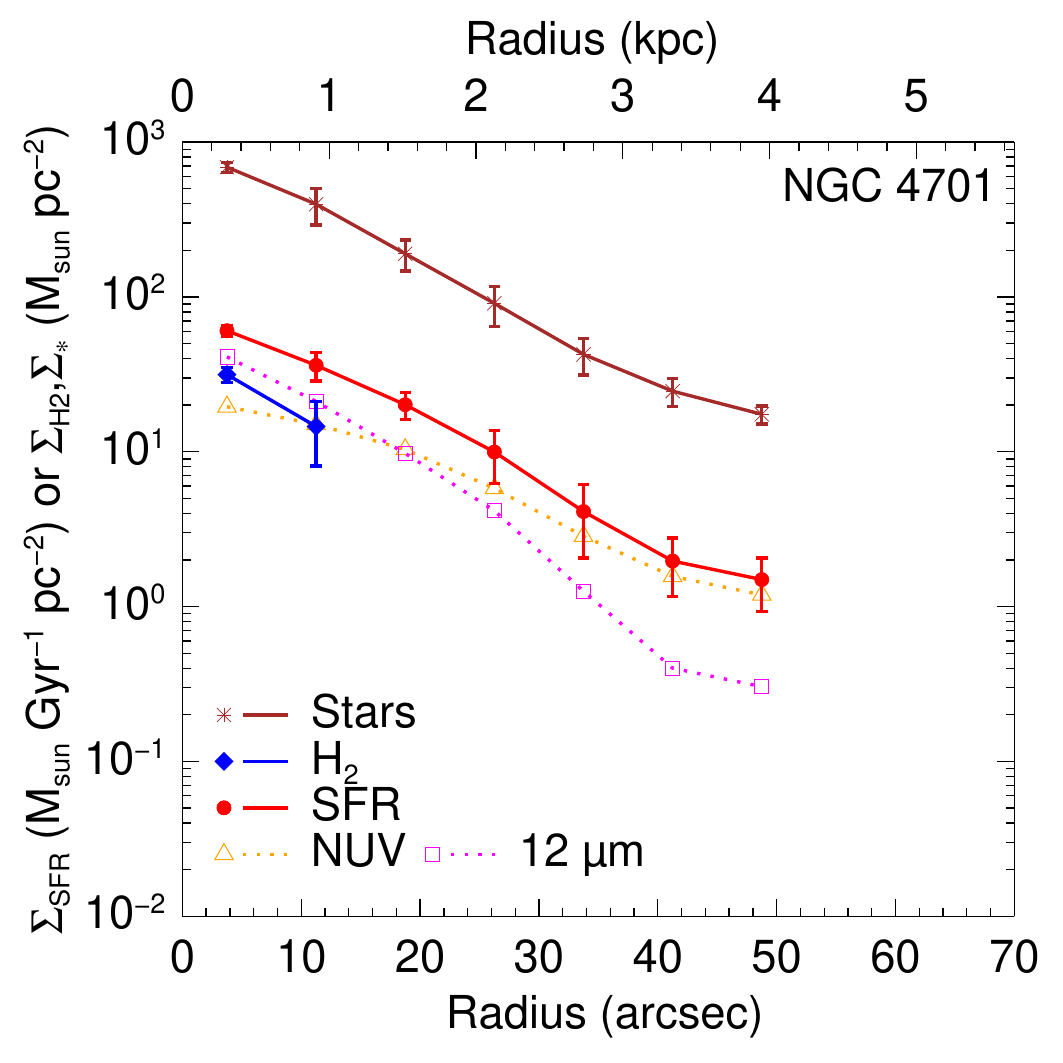}
\includegraphics[width=0.3\textwidth]{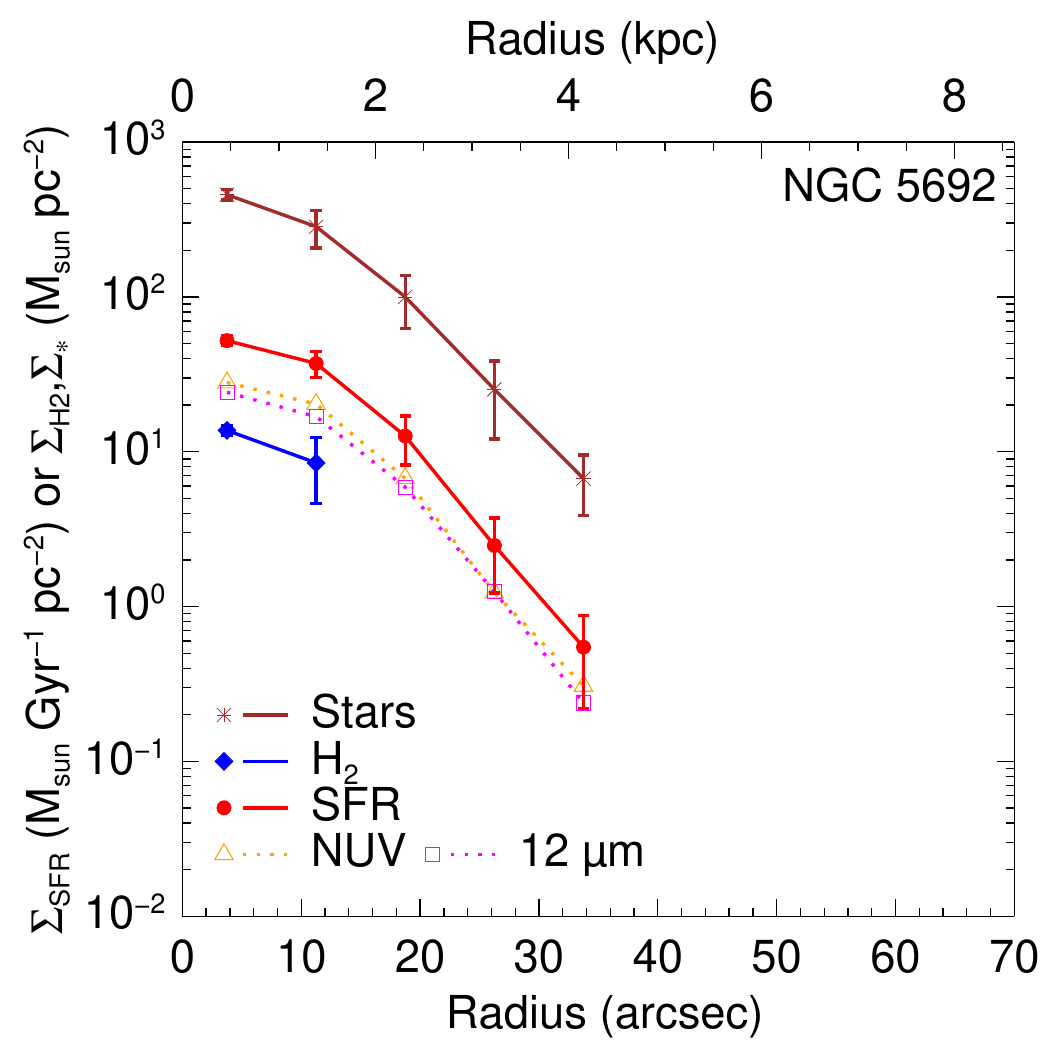}
\includegraphics[width=0.3\textwidth]{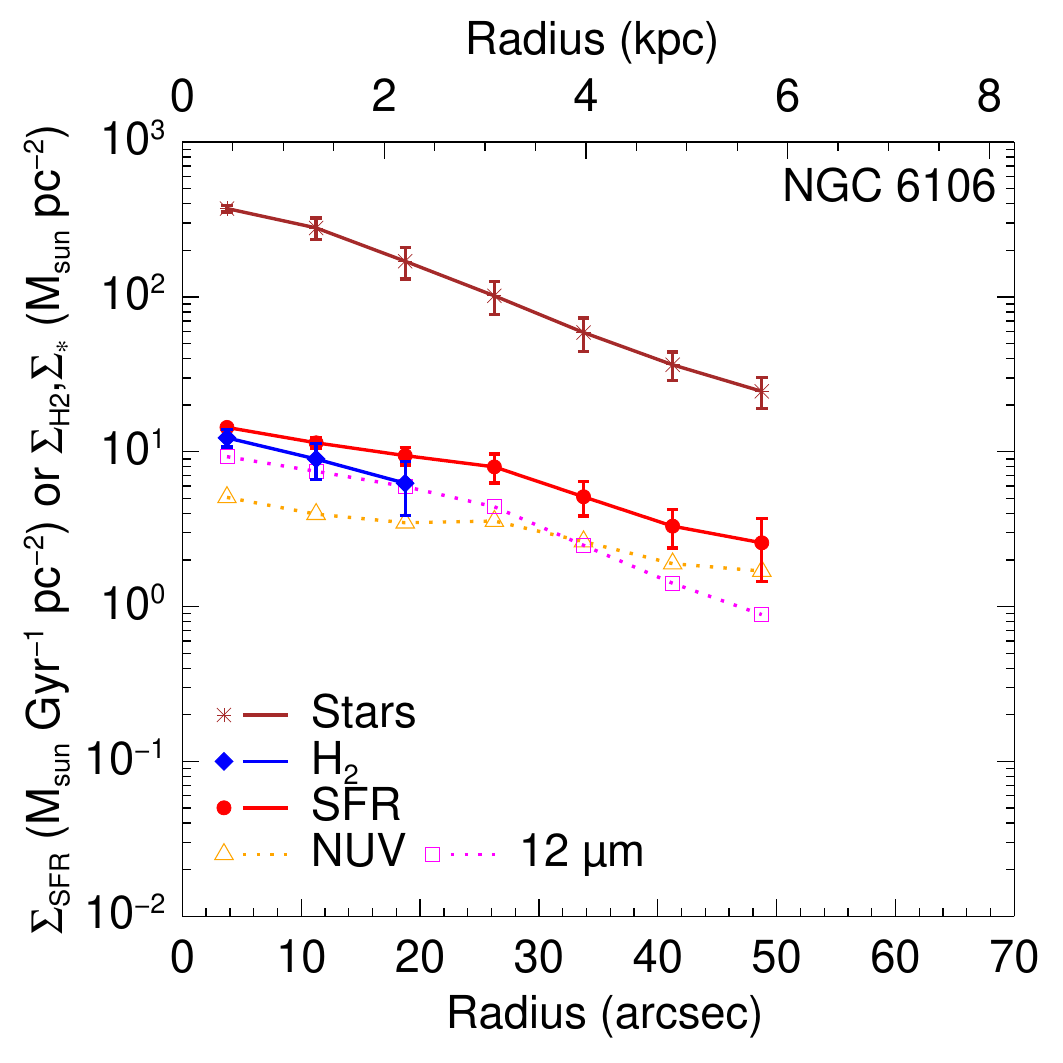}
\end{tabular}
\caption{Radial surface density profiles of \Htwo\ (blue diamonds), stars (brown stars), SFR (red circles), along with the individual contributions to the total SFR from NUV (orange triangles) and 12 \um\ (magenta squares). All profiles are measured in concentric elliptical annuli, each with a fixed radial width of 7.5\ac. The vertical error bars represent the standard deviation of data points in each annulus.
\label{fig:radialprofile}}
\end{center}
\end{figure*}

Figure \ref{fig:radialprofile} shows the radial profiles of \sightwo\ (blue diamonds), \sigsfr\ (red circles), \sigstar\ (brown stars), along with the individual contributions from NUV (orange triangles) and 12 \um\ (magenta squares) to the total SFR. We used the GIPSY task \texttt{ELLINT} to obtain the profiles measured in concentric elliptical annuli with a fixed radial width of 7.5\ac, which is consistent with the common spatial resolution of the convolved maps. The error bars denote the standard deviation of data points within each radial annulus. Overall, the profiles exhibit a decline with increasing radius. Most galaxies show a notable central concentration of molecular gas (\sightwo) within the inner 1--3 kpc. 
In contrast, SFR profiles are more spatially extended, indicating that recent star formation in dwarf galaxies can also occur at larger radii, even in regions where little or no molecular gas is detected. 

\subsection{The Molecular Star Formation Law}\label{sec:sflaw}

\begin{figure*}[!tbp]
\begin{center}
\begin{tabular}{c@{\hspace{0.1in}}c}
\includegraphics[width=0.5\textwidth]{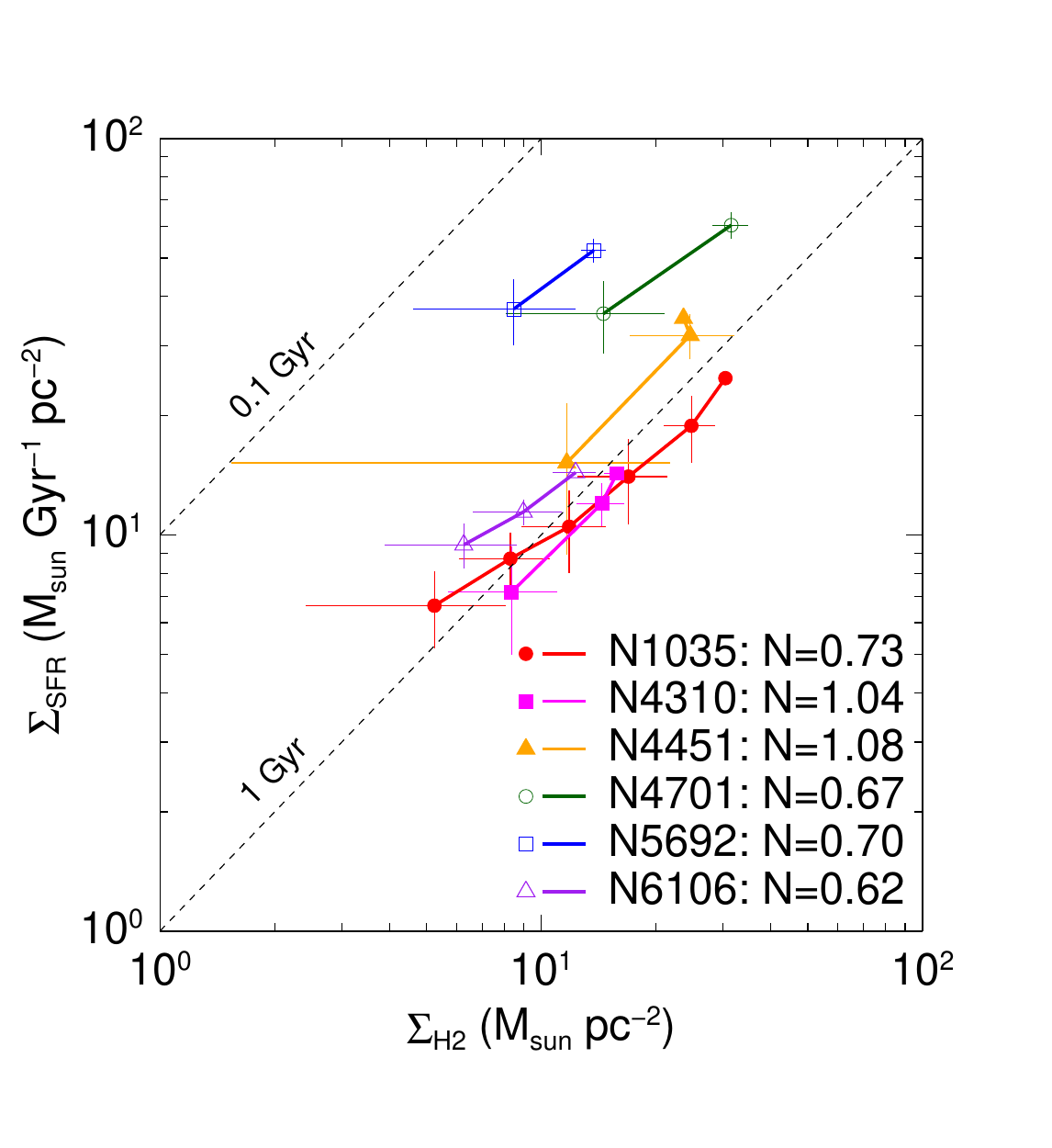} 
\includegraphics[width=0.5\textwidth]{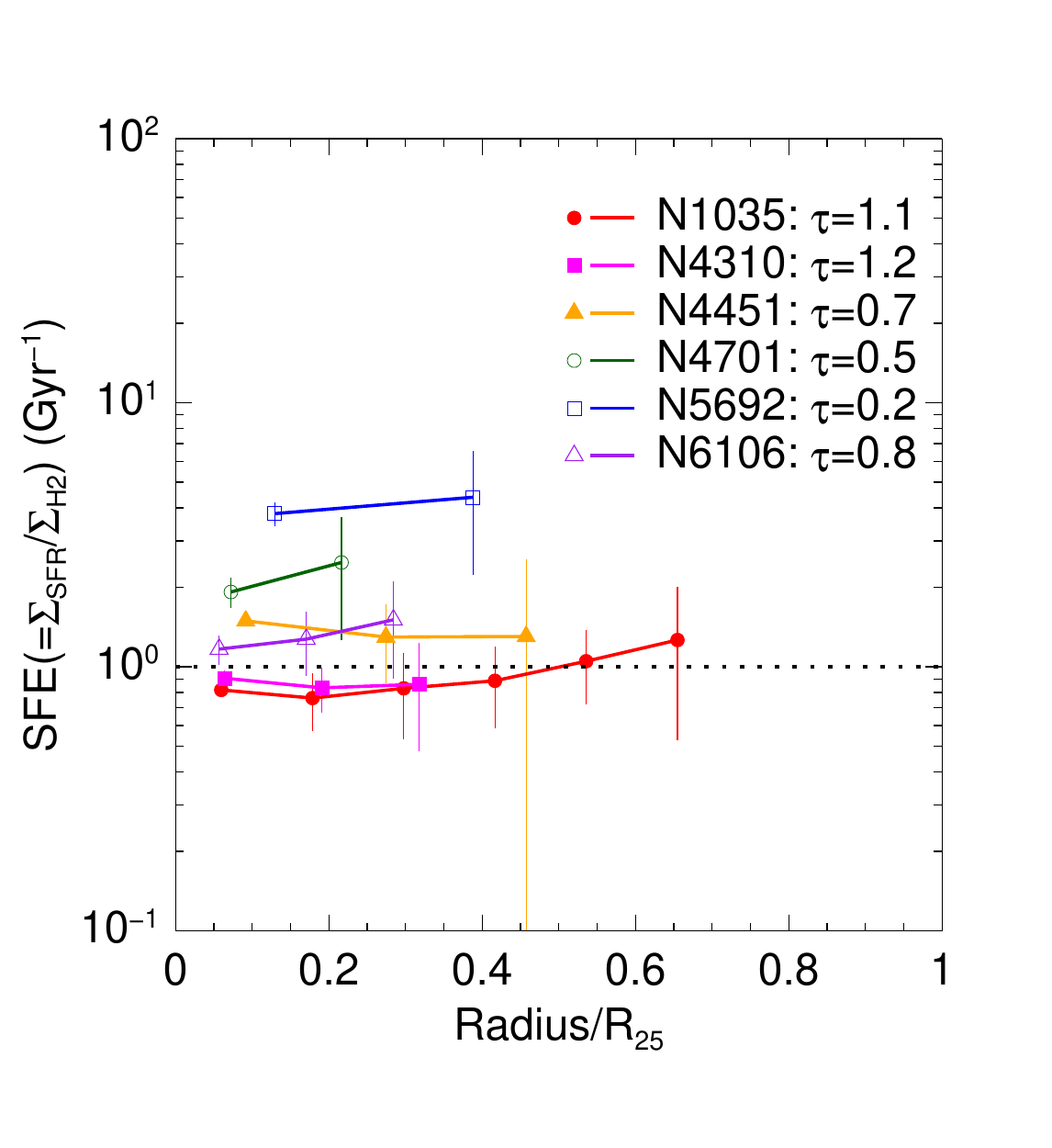} 
\end{tabular}
\caption{(Left) Relationship between \sigsfr\ and \sightwo\ for the six dwarf galaxies. The dashed diagonal lines represent constant molecular gas depletion times ($\tau_{H_2}$) of 0.1 and 1 Gyr for reference. The best-fit power-law index (N) is shown in the lower-right corner.  (Right) SFE of the molecular gas (SFE$_{\rm H_2}$) as a function of radius normalized by the optical radius (R$_{25}$). The molecular gas depletion time ($\tau_{H_2}$), measured in Gyr, is indicated for each galaxy in the upper-right corner. The horizontal dotted line marks  $\tau_{H_2}$ = 1 Gyr as a reference point. 
\label{kslaw}}
\end{center}
\end{figure*}

The SFL, often referred to as the KS relation, describes a power-law correlation between the surface densities of SFR and gas content (\HI\ + \Htwo): $\sigsfr \propto \siggas^{\rm N}$.
This relation quantifies the conversion of gas into stars in galaxies, where the slope N is an important parameter in galaxy evolution models and simulations  (e.g., \citealt{2010ApJ...721..975O}; \citealt{2012ApJ...745...69K}). 
In massive spiral galaxies, numerous studies have reported a nearly linear relationship between \sigsfr\ and \sightwo, with typical values of N ranging from 0.8 to 1.2 (e.g., \citealt{2002ApJ...569..157W}; \citealt{2008AJ....136.2846B}; \citealt{Leroy_2013}; \citealt{Sun_2023}). However, the applicability of this relation to low-mass, low-metallicity dwarf galaxies remains a topic of debate.

We examine the molecular SFL in our sample of dwarf galaxies using spatially resolved radial profiles of \sigsfr\ and \sightwo. In Figure \ref{kslaw} (left panel), we present the radial relation between \sigsfr\ and \sightwo, where we performed least-squares fits for each galaxy. The resulting power-law indices range from 0.62 to 1.08, with a mean slope of N = 0.81$\pm$0.18. This value is broadly consistent with the molecular SFL slope found in spiral galaxies (e.g., \citealt{2002ApJ...569..157W}; \citealt{2011AJ....142...37S}; \citealt{2020MNRAS.494.4558Y}).

For two galaxies in our sample (NGC 4701 and NGC 5692), the slope estimates rely on only two central data points where CO is significantly detected, which may lead to poorly constrained or biased results. Restricting the analysis to the four galaxies with three or more valid radial data points (NGC 1035, NGC 4310, NGC 4451, and NGC 6106) yields a mean slope of N = 0.87$\pm$0.05. This result is in good agreement with the slope (N$\approx$0.92) reported by the PHANGS-ALMA survey for nearby star-forming galaxies \citep{Sun_2023}.
Thus, our results suggest that when high-resolution and high-sensitivity ALMA CO data are used, the molecular SFL in dwarf galaxies does not differ significantly from that in more massive spirals.

To further assess whether the molecular SFL varies in the faint outer regions, we included additional data points down to the $1\sigma$ level of the CO intensity, except for NGC 4451, whose outermost point falls below $1\sigma$. Although these $1\sigma$ data are not considered significant detections, they serve as a useful test of the robustness of the relation. Including the $1\sigma$ data, the derived slopes only change slightly, yielding an average index of N$=0.73\pm 0.05$. The small variations imply that the molecular SFL inferred from the CO-bright central regions is broadly representative of the overall disk, and that the inclusion of the faintest CO-emitting regions does not significantly alter the relation.

We note that the adopted $\alpha_{\rm CO}$ introduces an additional systematic uncertainty. While $\alpha_{\rm CO}$ is expected to depend on local ISM properties such as metallicity (e.g., \citealt{2013ARA&A..51..207B}), spatially resolved metallicity measurements are not available for our sample. Moreover, the exact functional form of the metallicity-dependent $\alpha_{\rm CO}$ relation remains uncertain, and several previous studies adopted the Galactic value (e.g., \citealt{2008AJ....136.2782L}; \citealt{2011AJ....142...37S}).To assess the potential impact of a radially varying $\alpha_{\rm CO}$, we applied a simple, empirically motivated prescription based on Figure 7 of \citet{Sandstrom_2013}. Specifically, we reduced $\alpha_{\rm CO}$ to 70\% of the Galactic value in the central radial bin, adopted 90\% in the next radial bin, and retained the Galactic value beyond 0.3 R$_{25}$. Re-fitting the molecular SFL with this radially varying $\alpha_{\rm CO}$ prescription yields a modestly steeper slope of $\sim$1.1, compared to our result based on the constant $\alpha_{\rm CO}$. Importantly, this resulting slope still lies within the typical range found in spiral galaxies (N$\sim$0.8--1.2; \citealt{2002ApJ...569..157W}; \citealt{2008AJ....136.2846B}; \citealt{Leroy_2013}; \citealt{Sun_2023}) and does not alter our main conclusion that the molecular SFL in dwarf galaxies is broadly consistent with that of more massive spiral galaxies.



It is well established that gas volume density is more directly related to the gravitational collapse of molecular clouds and, consequently, to the star formation process than the surface density (e.g., \citealt{1959ApJ...129..243S}; \citealt{2012ApJ...745...69K}). However, measuring volume densities observationally requires knowledge of the disk thickness, which is often unavailable for dwarf galaxies. Recent studies by \citeauthor{2020MNRAS.494.4558Y} (\citeyear{2020MNRAS.494.4558Y}, \citeyear{Yim_2022})
have shown that although the power-law indices of the SFL differ significantly between volume and surface densities for the total gas (\HI\ + \Htwo), they remain similar when considering only the molecular gas component. Therefore, using surface densities to examine the molecular SFL, as adopted in this work, is both practical and physically well justified.

Figure \ref{kslaw} (right panel) shows the radial profiles of star formation efficiency (SFE) for the molecular gas, defined as SFE$_{\rm H_2} = \sigsfr/\sightwo$. This quantity reflects the efficiency with which molecular gas is converted into stars, and its inverse corresponds to the molecular gas depletion time, $\tau_{H_2}$. Understanding the radial behavior of SFE$_{\rm H_2}$ provides insight into the local regulation of star formation (\citealt{2008AJ....136.2782L}; \citealt{2010ApJ...721..975O}). 
In most galaxies, the SFE$_{\rm H_2}$ profiles are relatively flat across radius, showing only mild variations within R/R$_{25} \lesssim 0.8$. This radial uniformity is consistent with results from spiral galaxies, where the SFE has been found to remain nearly constant across disks (e.g., \citealt{1999AJ....118..670R}; \citealt{2008AJ....136.2846B}; \citealt{Leroy_2013}; \citealt{2016MNRAS.463.2092Y}). This suggests that where CO is reliably detected, molecular gas forms stars with similar efficiency regardless of galactocentric radius. 

In our sample, the SFE$_{\rm H_2}$ values range from 0.9 Gyr$^{-1}$ to 4.1 Gyr$^{-1}$, corresponding to depletion times of $\tau_{H_2} \sim$ 0.2--1.1 Gyr. These values are systematically shorter than the typical depletion time of $\sim$2 Gyr, often found in massive spirals
(\citeauthor{2008AJ....136.2846B} \citeyear{2008AJ....136.2846B}, \citeyear{Bigiel_2011};
\citealt{2008AJ....136.2782L}; \citealt{2011AJ....142...37S}; \citealt{2022AJ....164...43S}). 
However, they are consistent with studies of low-mass and low-metallicity systems (\citeauthor{2011AJ....142...37S} \citeyear{2011AJ....142...37S}, \citeyear{2012AJ....143..138S};
\citealt{Leroy_2013}; \citealt{2016A&A...588A..23A}). 
For example, \citet{2011AJ....142...37S} reported depletion times of $\sim$0.8 Gyr for dwarf galaxies with $M_* \leq 10^{10} M_\odot$ and 12 + log O/H $<$ 8.65. Additionally, \citet{2012AJ....143..138S} found that the depletion times of dwarf galaxies are one to two orders of magnitude shorter than those of massive spiral galaxies. \citet{2016A&A...588A..23A} measured $\tau_{H_2}$ $\sim$ 0.02--1 Gyr in 21 blue compact dwarfs, depending on their metallicities.     

The relatively short $\tau_{H_2}$ in our sample indicates that the molecular gas is consumed more rapidly in dwarf galaxies than in massive spirals, which may suggest enhanced SFE in these low-mass galaxies. 
However, several studies have argued that the short depletion times in low-metallicity dwarf galaxies are likely artificial, arising from the metallicity dependence $\alpha_{\rm CO}$. 
At low metallicity, $\alpha_{\rm CO}$ increases substantially because CO molecules are more easily photodissociated in dust-poor environments (\citealt{2011AJ....142...37S}, \citeyear{2012AJ....143..138S}; \citealt{2013ARA&A..51..207B}; \citealt{2016A&A...588A..23A}).
Consequently, applying a constant Galactic factor $\alpha_{\rm CO}$ underestimates the true \Htwo\ mass, leading to an artificially shortened depletion time (e.g., \citealt{2008AJ....136.2782L}; \citealt{2016A&A...588A..23A}). 

\subsection{Gravitational Instability}\label{sec:gravi}
To investigate how star formation is dynamically regulated in our sample of dwarf galaxies,  we examine the gravitational instability of their disks using the Toomre $Q$ parameter \citep{1964ApJ...139.1217T}. A lower $Q$ value indicates that the disk is more susceptible to gravitational collapse and becomes unstable when $Q < 1$ (\citealt{2001ApJ...555..301M}; \citealt{2013MNRAS.434.3389Z}). 

The Toomre instability parameters for gas and stars in a single-component rotating thin disk are defined according to \citet{1964ApJ...139.1217T} and \citet{2001MNRAS.323..445R}, respectively.
\begin{equation}
Q_{\rm g} = \frac{\kappa \sigma_{\rm g}}{\pi G \siggas}, \qquad 
Q_{\star} = \frac{\kappa \sigma_{\star,r}}{\pi G \sigstar},\\
\end{equation}
where
\begin{equation}
\kappa = \frac{V_{\rm rot}}{r} \sqrt{2\left(1 + \frac{r}{V_{\rm rot}}\frac{dV_{\rm rot}}{dr}\right)}.
\end{equation}
Here $\sigma_g$ is the gas velocity dispersion,  $\sigma_{\star,r}$ is the radial velocity dispersion of stars, $\kappa$ is the epicyclic frequency, and $V_{\rm rot}$ is the rotational velocity obtained from \citet{2022MNRAS.512.1012C}.

We adopt $\sigma_g = 5$ \kms, a representative value for dwarf galaxies \citep{2012AJ....143....1E}, and $\sigma_{\star,r} = 30$ \kms, which is consistent with expectations from the empirical stellar mass–dispersion relation \citep{2019MNRAS.489.3797M}. 
Since our sample consists of dwarf spiral galaxies, the adopted value of $\sigma_{\star,r}$ agrees well with values observed in late-type spirals, which typically range from 20 to 35 \kms\ \citep{2001MNRAS.323..445R}. The surface densities of gas (\siggas) and stars (\sigstar) are obtained from the radial profiles shown in Figure \ref{fig:radialprofile}.
Since \HI\ contributes little to the total surface mass density in the central regions of galaxies, where the molecular gas dominates the total gas, we compute $Q_{\rm g}$ using only \sightwo. Previous studies (e.g., \citealt{2011AJ....141...48Y}, \citeyear{2014AJ....148..127Y})
support this assumption, as they found that including \HI\ has a negligible effect on the total $Q$ value in the inner disk. 

\begin{figure}[!tbp]
\begin{center}
\includegraphics[width=0.5\textwidth]{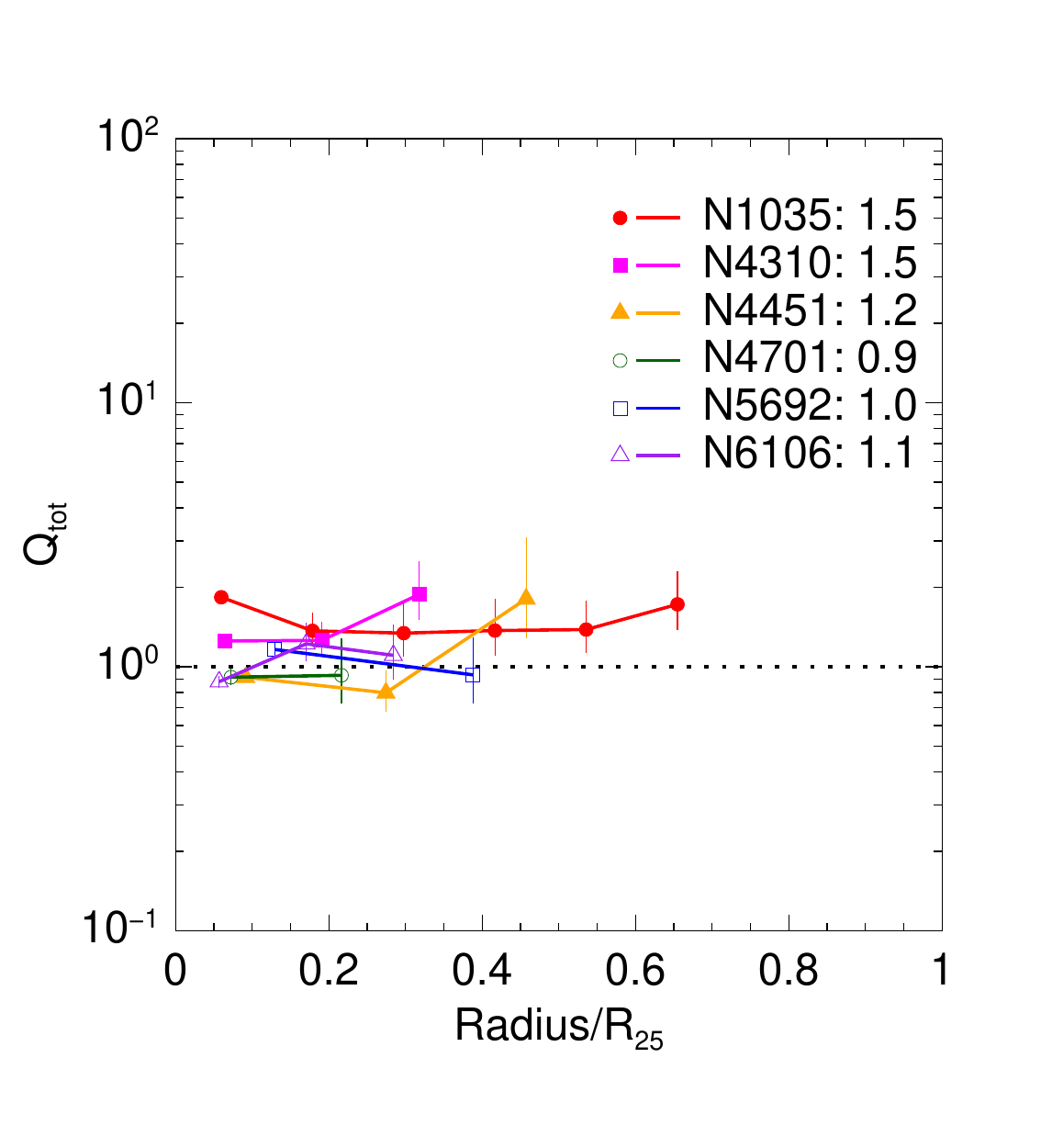} 
\caption{$Q_{\rm tot}$ as a function of radius, normalized by the optical radius $R_{25}$ for the six dwarf galaxies. The horizontal line at $Q_{\rm tot} = 1$ marks the threshold for gravitational instability; regions below the line are dynamically unstable and susceptible to star formation.
\label{fig:Qtot}}
\end{center}
\end{figure}

The total $Q$ parameter ($Q_{\rm tot}$) is calculated using the approximation given by \citet{2011MNRAS.416.1191R}, which accounts for both gas and stars in a thick disk:
\begin{equation}
\frac{1}{Q_{\rm tot}} = \left\{
\begin{array}{ll} 
\frac{\textrm {$W$}}{\textrm{$T_\star Q_{\star}$}} + \frac{\textrm{1}}{\textrm{$T_{\rm g}Q_{\rm g}$}} & \textrm{if\,\, $T_\star Q_{\star} \ge T_{\rm g}Q_{\rm g}$},\\
\\
\frac{\textrm{1}}{\textrm{$T_\star Q_{\star}$}} + \frac{\textrm{$W$}}{\textrm{$T_{\rm  g}Q_{\rm g}$}} & \textrm{ if\,\, $T_{\rm g}Q_{\rm g} \ge T_\star Q_{\star}$}, 
\end{array} \right.
\label{eq:Qtot}
\end{equation}
where
\begin{displaymath}
W = \frac{2s}{1+s^2}, \qquad  s = \frac{\sigma_{\rm g}}{\sigma_{\star,r}}, \qquad
T \approx 0.8 + 0.7 \left( \frac{\sigma_z}{\sigma_r} \right). 
\end{displaymath}
Here, $W$ is a weighting factor depending on the gas-to-stellar velocity dispersion ratio ($s$). The ratio of vertical to radial velocity dispersion ($\sigma_z/\sigma_r$) describes the dynamical anisotropy of each disk component and is used to compute the factor $T$, which quantifies the stabilizing effect of disk thickness in the two-component systems.
We adopt $\sigma_{z}/\sigma_{r}=0.6$ for stars, consistent with empirical and theoretical studies in dwarf galaxies (e.g., \citealt{2008AJ....136.2782L}; \citealt{2012AJ....143....1E}; \citealt{2020MNRAS.495.2867P}). For the gaseous component, we assume $\sigma_z/\sigma_r=1$, i.e., isotropic velocity dispersions, which is a reasonable approximation for cold gas disks (e.g., \citealt{2017MNRAS.469..286R}). 
Figure \ref{fig:Qtot} shows $Q_{\rm tot}$ as a function of radius, normalized by the optical radius $R_{25}$, for the six galaxies in our sample. The $Q_{\rm tot}$ values range from 0.9 to 1.5, with most galaxies remaining close to unity across their disks. Notably, the variation in $Q_{\rm tot}$ with radius is modest, showing no significant systematic increase or decrease within the CO-detected region ($R/R_{25} \lesssim 0.8$). This result suggests that the gravitational instability of the disks remains relatively uniform, implying marginal stability throughout the disk. This is consistent with results for massive spirals, where star-forming disks also remain near marginal stability over wide radial ranges (e.g., \citealt{2008AJ....136.2782L}; \citealt{2014AJ....148..127Y}).

Such marginally stable disks can be explained by self-regulated star formation (\citealt{1972ApJ...176L...9Q}; \citealt{2010ApJ...721..975O}; \citealt{2011ApJ...743...25K}; \citealt{2013MNRAS.433.1970F}). In this model, star formation is triggered locally when the disk becomes gravitationally unstable (i.e., when $Q < 1$). Feedback from newly formed stars--such as supernova explosions and stellar winds--increases the gas velocity dispersion and decreases the gas surface density, thereby raising the $Q$ parameter back toward unity. Conversely, when the disk is overly stable ($Q > 1$), star formation declines, allowing gas to accumulate, which eventually lowers Q and restores the conditions for instability. This negative feedback loop keeps the disk in a quasi-equilibrium state, where gravitational collapse and turbulent support are approximately balanced (\citealt{2011MNRAS.417..950H}; \citealt{2013MNRAS.434.3389Z}; \citealt{2017MNRAS.465.1682H}). Star formation thus proceeds in cyclic episodes, governed by the local interplay between instability-driven collapse and feedback-regulated suppression. 
Our observed radial profiles of $Q_{\rm tot}$, which are close to unity with little radial variation, are consistent with these theoretical predictions and support the idea that dwarf galaxy disks are dynamically self-regulated systems.

We note that our calculations assume constant velocity dispersions for both the gas and stellar components across the disk. However, observational studies suggest that these dispersions decrease with radius, particularly in massive spiral galaxies (\citealt{2009AJ....137.4424T}; \citealt{2010A&A...515A..62O}; \citealt{2014AJ....148..127Y}). In our case, the compact sizes of the dwarf disks and their relatively shallow potential wells likely minimize such radial variation. 

\section{Summary and Conclusions}
\label{sum}
In this study, we conducted a comprehensive analysis of molecular gas and star formation in six dwarf spiral galaxies. We used high-resolution archival ALMA \coj\ data along with multi-wavelength imaging data from GALEX (NUV), $Spitzer$ (3.6 \um), and WISE (3.4 \um\ and 12 \um). The primary goal of this work is to examine whether dwarf galaxies, despite their low mass and metallicity, follow the molecular SFL observed in massive spirals and have their star formation regulated by gravitational stability as quantified by the Toomre $Q$ parameter. This question has remained unresolved due to previous observational limitations, particularly the challenge of detecting faint CO emissions in dwarf galaxies. By taking advantage of the superior spatial resolution and sensitivity of ALMA, we provide a more reliable assessment of the molecular gas–SFR relation, as well as the role of gravitational instability in governing star formation activity in these low-mass galaxies.
\\

1. We examined the spatially resolved SFL for each galaxy based on radial surface density profiles of the SFR and molecular gas. The fitted power-law indices range from 0.62 to 1.08, with a sample-averaged slope of N = $0.81\pm0.18$. Excluding galaxies with only two CO-detected radial bins, the average slope increases to N = $0.87\pm0.05$, consistent with the range reported for nearby star-forming spirals (e.g., \citealt{Sun_2023}). This result suggests that, on scales of several hundred parsecs, dwarf galaxies can exhibit molecular SFLs comparable to those of massive galaxies.

2. The molecular gas depletion times in our sample range from 0.2 to 1.1 Gyr, which are systematically shorter than the typical depletion time of $\sim$2 Gyr found in massive spiral galaxies. However, these short timescales are likely a consequence of adopting the standard Galactic conversion factor, $\alpha_{\rm CO} = 4.35\ M_\odot\ \rm pc^{-2}\ [K \,\kms]^{-1}$ , rather than evidence for enhanced star formation efficiency. The relatively flat radial profiles of the molecular SFE further support the idea that molecular gas forms stars with comparable efficiency across the disk where CO is detected.

3. We computed the gravitational instability parameter $Q_{\rm tot}$ using a two-component formulation that includes both stars and gas. The $Q_{\rm tot}$ values remain close to unity across most of the disks, with minimal radial variation, as reported in massive spiral galaxies. This state of marginal stability is consistent with self-regulated star formation models, in which stellar feedback counteracts gravitational collapse to maintain a quasi-equilibrium.

Our results demonstrate that, despite differences in mass and chemical abundance, the molecular SFL and gravitational instability in dwarf galaxies behave similarly to those of more massive systems. 
This highlights the importance of high-resolution, high-sensitivity molecular gas observations for establishing continuity in star formation physics across a broad range of galaxy types. 
To better test the generality of our results, future work would benefit from enlarging the sample to include more dwarf galaxies with population diversity (e.g., variations in metallicity, morphology, and environment) using high-quality CO observations. 
Upcoming observations with ALMA, together with next-generation facilities like the ngVLA and SKA, will be essential for overcoming current sample limitations and further testing the universality of the molecular SFL and gravitational stability across galaxy populations.

\begin{acknowledgments}
We thank the anonymous referee for useful suggestions that improved this paper.
This work was supported by the National Research Foundation of Korea through grants, NRF-2022R1A2C1007721.
This paper makes use of the following ALMA data: ADS/JAO.ALMA\#2015.1.00820.S. ALMA is a partnership of ESO (representing its member states), NSF (USA) and NINS (Japan), together with NRC (Canada), NSTC and ASIAA (Taiwan), and KASI (Republic of Korea), in cooperation with the Republic of Chile. The Joint ALMA Observatory is operated by ESO, AUI/NRAO and NAOJ.
\end{acknowledgments}





%
\facilities{ALMA, $Spitzer$, WISE, GALEX}




\bibliography{refer}{}
\bibliographystyle{aasjournalv7}



\end{document}